\documentclass{llncs}
\usepackage[english]{babel}
\usepackage[utf8]{inputenc}
\usepackage{amsmath}
\usepackage{amsfonts}
\usepackage{amssymb}
\usepackage{graphicx}
\usepackage[ruled,noline,noend]{algorithm2e}
\usepackage{multirow}
\usepackage{tabulary}

\begin{document}

\title{Improving incremental recommenders with online bagging}

\author{João Vinagre\inst{1} \and Alípio Mário Jorge\inst{1,2} \and João Gama\inst{1,3}}

\institute{
LIAAD - INESC TEC, Porto, Portugal\\
\and
Faculty of Sciences, University of Porto, Portugal\\
\and
Faculty of Economics, University of Porto, Portugal\\
\email{jnsilva@inesctec.pt}, \email{amjorge@fc.up.pt}, \email{jgama@fep.up.pt}
}

\maketitle

\begin{abstract}
Online recommender systems often deal with continuous, potentially fast and unbounded flows of data. Ensemble methods for recommender systems have been used in the past in batch algorithms, however they have never been studied with incremental algorithms that learn from data streams. We evaluate online bagging with an incremental matrix factorization algorithm for top-$N$ recommendation with positive-only -- binary -- ratings. Our results show that online bagging is able to improve accuracy up to 35\% over the baseline, with small computational overhead.
\end{abstract}


\section{Introduction}

In many real world recommender systems, user feedback is continuously generated at unpredictable rates and order, and is potentially unbounded. In large scale systems, the rate at which user feedback is generated can be very fast. Building predictive models from these continuous flows of data is a problem actively studied in the field of data stream mining. Ideally, algorithms that learn from data streams should be able to process data at least as fast as it arrives, in a single pass, while maintaining an always-available model \cite{DBLP:conf/dmkd/DomingosH01}. Most incremental algorithms naturally have these properties, and are thus a viable solution.

Incremental algorithms for recommendation also treat user feedback data as a data stream, immediately incorporating new data in the recommendation model. In many -- if not most -- recommendation applications this is a desirable feature, since it gives the model the ability to evolve over time. This is important, because the task of a recommender system is to find the most relevant items to each user, individually. Naturally, users are human beings, whose preferences change over time. Moreover, in large scale systems, new users and items are permanently entering the system. A model that is immediately updated with fresh data has the capability of adjusting faster to such changes.

\subsection{Related work}

Ensemble methods in machine learning are convenient techniques to improve the accuracy of algorithms. Typically, this is achieved by combining results from a number of weaker sub-models. Bagging \cite{DBLP:journals/ml/Breiman96b}, Boosting \cite{DBLP:conf/icml/FreundS96} and Stacking \cite{DBLP:journals/nn/Wolpert92} are three well-known ensemble methods used with recommendation algorithms. Boosting is experimented in \cite{DBLP:conf/pkdd/ChowdhuryCL15,DBLP:conf/recsys/SchclarTRMA09,DBLP:conf/kdd/JahrerTL10,segrera2006experimental}, bagging is studied also in \cite{DBLP:conf/kdd/JahrerTL10,segrera2006experimental}, and stacking in \cite{DBLP:journals/corr/abs-0911-0460}. In all of these contributions, ensemble methods work with batch learning algorithms only.

In this paper we propose online bagging for incremental recommendation algorithms designed to deal with streams of positive user feedback. To our best knowledge this is the first ensemble method proposed for incremental recommender systems in the literature.

This paper is organized as follows. After this introductory section, we describe online bagging for common data mining tasks in Section \ref{sec:online_bagging}. In Section \ref{sec:eval_bagging} we present our experiments and results, along with a short discussion. Finally, we conclude in Section \ref{sec:conclusions}.

\section{Online bagging}
\label{sec:online_bagging}

Bagging \cite{DBLP:journals/ml/Breiman96b} is an ensemble technique that takes a number of bootstrap samples of a dataset and trains a model on each one of the samples. Predictions from the various sub-models are then aggregated in a final prediction. This is known to improve the performance of algorithms by reducing variance, which is especially useful with unstable algorithms that are very sensitive to small changes in the data. The diversity offered by training several models with slightly different bootstrap samples of the data helps in giving more importance to the main concepts being learned -- since they must be present in most bootstrap samples of the data -- , and less importance to noise or irrelevant phenomena that may mislead the learning algorithm.

To obtain a bootstrap sample of a dataset with size $N$, we perform $N$ trials, sampling a random example with replacement from the dataset. Each example has probability of $1/N$ to be sampled at each trial. The resulting dataset will have the same size of the original dataset, however some examples will not be present whereas some others will occur multiple times. To obtain $M$ samples, we simply repeat the process $M$ times.

In its original proposal \cite{DBLP:journals/ml/Breiman96b}, bagging is a batch procedure requiring $N \times M$ passes through the dataset. However, it has been shown in \cite{DBLP:conf/kdd/OzaR01} that this can be done incrementally in a single pass, if the number of examples is very large -- a natural assumption when learning from data streams. Looking at the batch method above, we observe that each bootstrap sample contains $K$ occurrences of each example, with $K \in \{0,1,2,\ldots\}$, and:

\begin{equation}
\label{eq:bagg1}
P(K = k) = {N \choose k} \left(\frac{1}{N}\right)^k \left(1 - \frac{1}{N}\right)^{N-k}
\end{equation}

In an incremental setting, one could just initialize $M$ sub-models -- or bootstrap nodes -- and then use \eqref{eq:bagg1} to train new examples $K$ times, redrawing $K$ for each node. The problem is that this would still require knowing $N$ beforehand. However, if we assume that $N \rightarrow \infty$, then the distribution of $K$ tends to a $Poisson(1)$ distribution, and therefore

\begin{equation}
\label{eq:bagg2}
P(K = k) = \frac{e^{-1}}{k!}
\end{equation}

eliminating the need of any prior knowledge about the data, allowing the usage of bagging in a single pass over data.

\section{Online recommendation with bagging}
\label{sec:online_bagging_rec}

To assess the potential of online bagging, we use ISGD \cite{DBLP:conf/um/VinagreJG14}, a simple online matrix factorization algorithm for positive-only data. ISGD (Algorithm \ref{alg:ISGD}) uses Stochastic Gradient Descent in one pass through the data, which is convenient for data stream processing. It is designed for positive-only streams of user-item pairs $(u,i)$ that indicate a positive interaction between user $u$ and item $i$. Examples of positive interactions are users buying items in an online store, streaming music tracks from an online music streaming service, or simply visiting web pages. This is a much more widely available form of user feedback, than for example, ratings data, which is only available from systems with user rating features.

\begin{algorithm}
\DontPrintSemicolon
\SetKwInOut{Input}{input}\SetKwInOut{Output}{output}
\SetKwFunction{Shuffle}{shuffle}
\SetKwFunction{Vector}{Vector}
\SetKwFunction{Rows}{Rows}
\SetKwData{Size}{size}
\KwData{a finite set or a data stream $D = \{(u,i)_1,(u,i)_2,\ldots\}$}
\Input{no. of latent features $k$, no. of iterations $iter$, regularization factor $\lambda$, learn rate $\eta$}
\Output{user and item factor matrices $A$ and $B$}
\BlankLine
\For{$(u,i) \in D$} {
	\If{$u \not \in \Rows(A)$} {
		$A_u \leftarrow \Vector(\Size:k)$\;
		$A_u \sim \mathcal{N}(0,0.1)$\;
	}
	\If{$i \not \in \Rows(B)$} {
		$B_i \leftarrow \Vector(\Size:k)$\;
		$B_i \sim \mathcal{N}(0,0.1)$\;
	}
	\For{$n \leftarrow 1$ \KwTo $iter$} {
		$err_{ui} \leftarrow 1 - A_u \cdot B_i$\;
		$A_u \leftarrow A_u + \eta (err_{ui} B_i - \lambda A_u)$\;
		$B_i \leftarrow B_i + \eta (err_{ui} A_u - \lambda B_i)$\;
	}
}
\caption{ISGD - Incremental SGD for positive-only ratings \cite{DBLP:conf/um/VinagreJG14}}
\label{alg:ISGD}
\end{algorithm} 

ISGD continuously updates factor matrices $A$ -- the user factors matrix -- and $B$ -- the item factors matrix --, correcting the model to adapt to the incoming user-item pairs. If $(u,i)$ occurs in the stream, then the model prediction $\hat{R}_{ui} = A_u \cdot B_i$ should be close to 1. Top-$N$ recommendations to any user $u$ is obtained by ranking function  $f = |1 - \hat{R}_{ui}|$ for all items $i$ in ascending order, and taking the top $N$ items.

The online bagging approach described in Section \ref{sec:online_bagging}, can be easily applied to ISGD, resulting in Algorithm \ref{alg:bagged_isgd} -- BaggedISGD.

\begin{algorithm}
\DontPrintSemicolon
\SetKwInOut{Input}{input}\SetKwInOut{Output}{output}
\SetKwFunction{Shuffle}{shuffle}
\SetKwFunction{Vector}{Vector}
\SetKwFunction{Rows}{Rows}
\SetKwData{Size}{size}
\KwData{a finite set or a data stream of user-item pairs $D = \{(u,i)_1,(u,i)_2,\ldots\}$}
\Input{no. of latent features $k$, no. of iterations $iter$, regularization factor $\lambda$, learn rate $\eta$, no. of bootstrap nodes $M$}
\Output{$M$ user and item factor matrices $A^m$ and $B^m$}
\BlankLine
\For{$(u,i) \in D$} {
	\For{$m \leftarrow 1$ \KwTo $M$} {
		$k \sim \operatorname{Poisson}(1)$ \tcp*{eq. \eqref{eq:bagg2}}
		\If{$k > 0$} {
			\For{$l \leftarrow 1$ \KwTo $k$} {
				\If{$u \not \in \Rows(A^m)$} {
					$A^m_u \leftarrow \Vector(\Size:k)$\;
					$A^m_u \sim \mathcal{N}(0,0.1)$\;
				}
				\If{$i \not \in \Rows(B^m)$} {
					$B^m_i \leftarrow \Vector(\Size:k)$\;
					$B^m_i \sim \mathcal{N}(0,0.1)$\;
				}
				\For{$n \leftarrow 1$ \KwTo $iter$} {
					$err_{ui} \leftarrow 1 - A^m_u \cdot B^m_i$\;
					$A^m_u \leftarrow A^m_u + \eta (err_{ui} B^m_i - \lambda A^m_u)$\;
					$B^m_i \leftarrow B^m_i + \eta (err_{ui} A^m_u - \lambda B^m_i)$\;
				}
			}
		}
	}
}
\caption{BaggedISGD - Bagging version of ISGD (training algorithm)}
\label{alg:bagged_isgd}
\end{algorithm}

BaggedISGD learns $M$ models on $M$ bootstrap nodes, each of them based on the online bootstrap sampling method described in Section \ref{sec:online_bagging}. Similarly to ISGD, to perform the actual list of recommendations for a user $u$, items $i$ are sorted by a function $f = |1 - \hat{R}_{ui}|$. However, the scores $\hat{R}_{ui}$ are actually the average score of all nodes:

\begin{equation}
\hat{R}_{ui} = \frac{\sum_{m = 1}^MA_u^m \cdot B_i^m}{M}
\end{equation}

At training time, this algorithm requires at least $M$ times the computational resources needed for ISGD, with $M$ bootstrap nodes. Recommendation also has the overhead of aggregating $M$ predictions from the submodels. In our experiments, we also measure update and recommendation times, for several values of $M$.

\section{Evaluation}
\label{sec:eval_bagging}

To simulate a streaming environment we need datasets that maintain the natural order of the data points, as they were generated. Additionally, we need positive-only data, since the tested algorithm is not designed to deal with ratings. We use 4 datasets that conciliate these two requirements -- positive-only and naturally ordered --, described in Table \ref{tab:datasets}. ML1M is based on the Movielens-1M movie rating dataset\footnote{\url{http://www.grouplens.org/data} [Jan 2013]}. To obtain the YHM-6KU, we sample 6000 users randomly from the Yahoo! Music dataset\footnote{\url{https://webscope.sandbox.yahoo.com/catalog.php?datatype=r} [Jan 2013]}. LFM-50U is a subset consisting of a random sample of 50 users taken from the Last.fm\footnote{\url{http://last.fm/}} dataset\footnote{\url{http://ocelma.net/MusicRecommendationDataset} [Jan 2013]}. PLC-STR \footnote{\url{https://rdm.inesctec.pt/dataset/cs-2017-003}, file: \texttt{playlisted\_tracks.tsv}} consists of the music streaming history taken from Palco Principal\footnote{\url{http://www.palcoprincipal.com/}}, a portuguese social network for non-mainstream artists and fans. 

All of the 4 datasets consist of a chronologically ordered sequence of positive user-item interactions. However, ML1M and YHM-50U are obtained from ratings datasets. To use them as positive-only data, we retain the user-item pairs for which the rating is in the top 20\% of the rating scale. This means retaining only the rating 5 in ML1M and rating of 80 or more in the YHM-6KU dataset. Naturally, only single accurrences of user-item pairs are available in these datasets, since users do not rate the same item more than once. PLC-STR and LFM-50 have multiple occurrences of the same user-item pairs.

\begin{table}
\centering
\footnotesize
\begin{tabular}{l r r r r r}
\hline
Dataset & Events & Users & Items & Sparsity \\
\hline
PLC-STR & 588 851 & 7 580 & 30 092 & 99.74\% \\
\hline
LFM-50U & 1 121 520 & 50 & 159 208 & 85.91\% \\
\hline
YHM-6KU & 476 886 & 6 000 & 127 448 & 99.94\% \\
\hline
ML1M & 226 310 & 6 014 & 3 232 & 98.84\% \\
\hline
\end{tabular}
\caption{Dataset description}\normalsize
\label{tab:datasets}
\end{table}

We run a set of experiments using the prequential approach \cite{DBLP:journals/ml/GamaSR13} as described in \cite{DBLP:conf/um/VinagreJG14}. Each observation in the dataset consists of a simple user-item pair $(u,i)$ that indicates a positive interaction between user $u$ and item $i$. The following steps are performed in the prequential evaluation process:

\begin{enumerate}
\setlength\itemsep{0em}
  \item If $u$ is a known user, use the current model to recommend a list of items to $u$, otherwise go to step 3;
  \item Score the recommended list given the observed item $i$;
  \item Update the model with $(u,i)$ (optionally);
  \item Proceed to -- or wait for -- the next observation
\end{enumerate} 

This process is entirely applicable to algorithms that learn either incrementally or in batch mode. This is the reason why step 3. is annotated as optional. For example, instead of performing this step, the system can store the data to perform batch retraining periodically.
\begin{table}
\center
\begin{tabular}{l|l|r|r|r|r|r|r}
\hline
Dataset & $M$ & Rec@1 & Rec@5 & Rec@10 & Rec@20 & Upd. (ms) & Rec. (ms) \\
\hline
	\multirow{4}{*}{PLC-STR}
	& ISGD & \textbf{0.127} & \textbf{0.241} & 0.277 & 0.302 & 0.237 & 21.736 \\
	& 8 & 0.076 & 0.194 & 0.257 & 0.316 & 2.563 & 64.793 \\
	& 16 & 0.081 & 0.215 & 0.284 & 0.349 & 4.732 & 132.812 \\
	& 32 & 0.088 & 0.229 & 0.302 & 0.370 & 9.508 & 264.846 \\
	& 64 & 0.092 & 0.237 & \textbf{0.313} & \textbf{0.384} & 18.012 & 517.479 \\
\hline
	\multirow{4}{*}{LFM-50U}
	& ISGD & \textbf{0.034} & 0.049 & 0.052 & 0.055 & 2.625 & 94.177 \\
	& 8 & 0.023 & 0.044 & 0.052 & 0.058 & 21.449 & 241.452 \\
	& 16 & 0.026 & 0.050 & 0.059 & 0.066 & 43.094 & 491.689 \\
	& 32 & 0.028 & 0.055 & 0.064 & 0.071 & 84.536 & 984.060 \\
	& 64 & 0.030 & \textbf{0.057} & \textbf{0.067} & \textbf{0.075} & 168.781 & 1.958 s \\
\hline
	\multirow{4}{*}{YHM-6KU}
	& ISGD & \textbf{0.030} & \textbf{0.063} & 0.082 & 0.103 & 4.462 & 89.321 \\
	& 8 & 0.011 & 0.033 & 0.051 & 0.076 & 28.529 & 347.422 \\
	& 16 & 0.012 & 0.037 & 0.058 & 0.086 & 54.723 & 667.898 \\
	& 32 & 0.019 & 0.055 & 0.082 & 0.117 & 158.744 & 990.551 \\
	& 64 & 0.021 & 0.059 & \textbf{0.087} & \textbf{0.123} & 328.924 & 1.934 s \\
\hline
	\multirow{4}{*}{ML1M}
	& ISGD & 0.005 & 0.021 & 0.034 & 0.055 & 0.069 & 2.557 \\
	& 8 & 0.005 & 0.019 & 0.033 & 0.056 & 0.517 & 7.208 \\
	& 16 & 0.006 & 0.022 & 0.038 & 0.063 & 1.390 & 21.816 \\
	& 32 & 0.006 & 0.025 & 0.042 & 0.071 & 1.866 & 33.496 \\
	& 64 & \textbf{0.007} & \textbf{0.026} & \textbf{0.045} & \textbf{0.074} & 3.999 & 41.090 \\
\hline
\end{tabular}
\caption[ISGD with bagging]{Average performance of ISGD with and without bagging. $M$ is the number of bootstrap nodes. The last two columns contain the average update times and the average recommendation times.}
\label{tab:results_bagging}
\end{table}

To kickstart the evaluation process we use 10\% the available data to train a base model in batch, and use the remaining 90\% to perform incremental training and evaluation. We do this initial batch training to avoid \emph{cold-start} problems, which are not the subject of our research. 

In our setting, the items that users have already co-occurred with -- i.e. items that users know -- are not recommended. This has one important implication in the prequential evaluation process, specifically on datasets that have multiple occurrences of the same user-item pair. Evaluation at these points is necessarily penalized, since the observed item will be not be within the recommendations. In such cases, we bypass the scoring step, but still use the observation to update the model.  

We measure two dimensions on the evaluation process: accuracy and time. In the prequential process described above, we need to make a prediction and evaluate it at every new user-item pair $(u,i)$ that arrives in the data stream. To do this, we use the current model to recommend a list of items to user $u$. We then score this recommendation list, by matching it to the actually observed item $i$. We use a recommendation list with at most 20 items, and then score this list as 1 if $i$ is within the  recommended items, and 0 otherwise, using Recall@$C$ with cutoffs $C \in \{1,5,10,20\}$. Because only one item is tested against the list, Recall@$C$ can only take the values $\{0,1\}$. We can calculate the overall Recall@$C$ by averaging the scores at every step. Additionally, we can also depict it using a moving average. Time is measured in milliseconds at every step and we depict it using the same techniques we use with accuracy.

All experiments were run in Intel Haswell 4-core machines, with CentOS Linux 7 64 bit. The algorithms and prequential evaluation code is implemented on top of MyMediaLite \cite{DBLP:conf/recsys/GantnerRFS11}. The recommendation step is implemented with multi-core code -- predictions from nodes are computed in parallel.

\subsection{Results}
\label{subsec:results}

To evaluate bagging, we experiment with four levels of bootstrapping $M \in \{8,16,32,64\}$. Table \ref{tab:results_bagging} summarizes the results of our experiments. Values in Table \ref{tab:results_bagging} are obtained by averaging Recall and time obtained at all prequential evaluation steps. With all datasets except YHM-6KU, bagging improves the Recall, especially with $M \geq 32$. One interesting observation is that bagging has a bigger influence on higher Recall cutoffs, which suggests that improvements of the predictive ability are typically not obtained in the top 5 recommended items. 

The model update times increase approximately in proportion to the number of bootstrap nodes $M$, which is not surprising, since the algorithm performs the update operations one time (in average) in each one of the $M$ bootstrap nodes. However, since the baseline update time is very small, this overhead is also small. The last column of Table \ref{tab:results_bagging} contains the recommendation time, specifically the average time required to produce a recommendation list. The bagging algorithm needs to aggregate predictions coming from all $M$ nodes, which is an important overhead. Results show that both the update times and recommendation times increase proportionally to $M$. However, the recommendation step is a far more expensive operation, even when computed in parallel. For example, using $M = 64$ with LFM-50U and YHM-6KU, recommendations are computed in nearly two seconds in average, in 4-core machines, which can reasonably be considered too much in many applications.

\begin{figure*}[ht!] 
\center
\includegraphics[width=\textwidth]{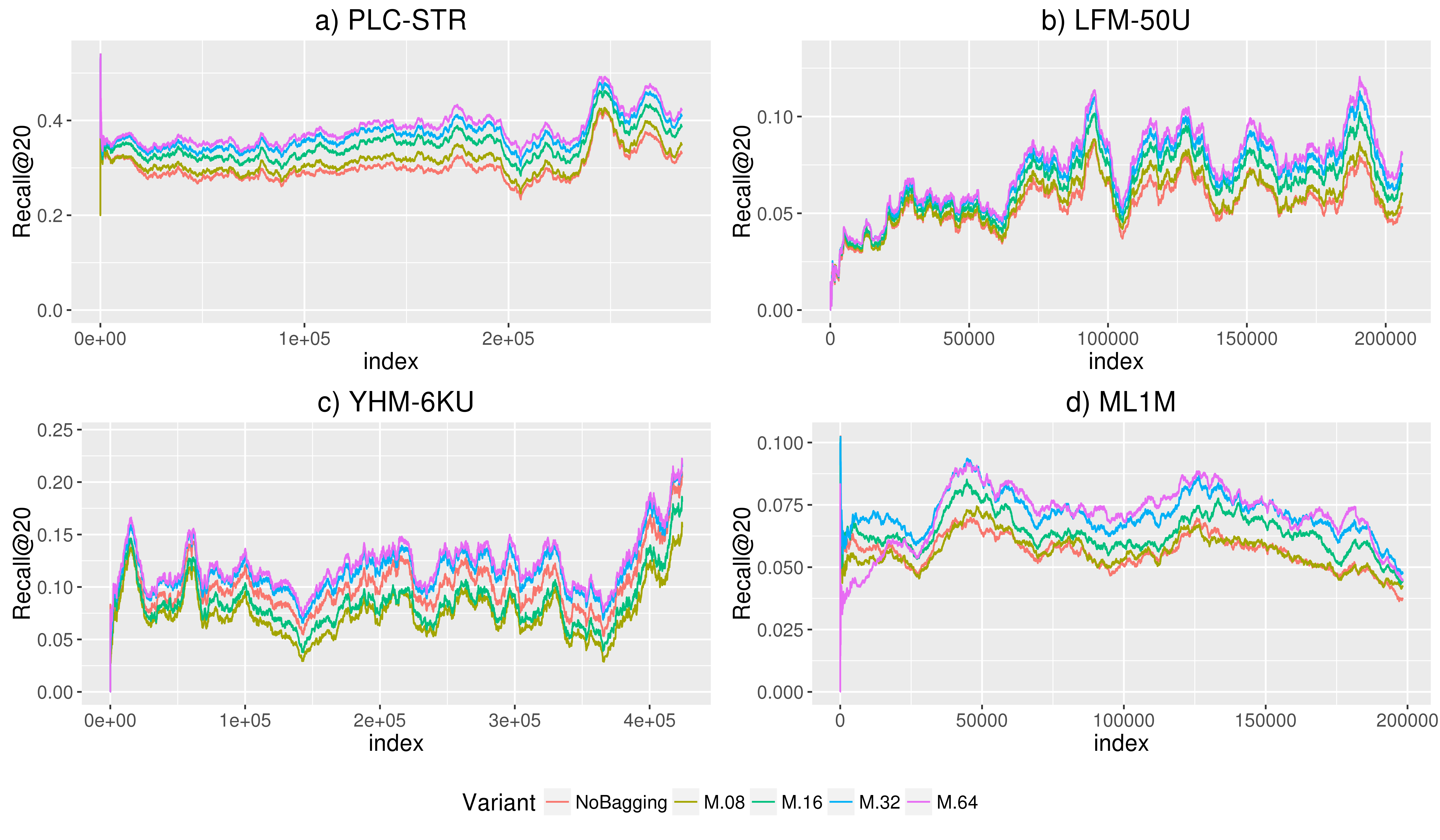}
\caption[Prequential evaluation of Recall@20 with ISGD using bagging]{Prequential evaluation of Recall@20 with ISGD with and without bagging. Lines are drawn using a moving average of Recall@20 with $n = 10 000$. The first 10 000 points are drawn using the accumulated average.}
\label{fig:recall20_bagging}
\end{figure*}

A useful feature of prequential evaluation is that it allows us also to depict the evolution of Recall@20 in Figure \ref{fig:recall20_bagging}. This visualization reveals how the predictive ability of the algorithm performs over time, as the incremental learning process occurs. For the sake of space, we omit other cutoffs.

\subsection{Discussion}

Results in Table \ref{tab:results_bagging} and Figure \ref{fig:recall20_bagging} show that bagging clearly improves the accuracy of ISGD, with accuracy improvements of  35\% over the baseline (see Table \ref{tab:results_bagging} LFM-50U and ML1M). This improvement is mainly observable with cutoffs $C \geq 5$ of Recall. Given that bagging  reduces variance \cite{DBLP:journals/ml/Breiman96b}, this suggests that the variance of ISGD is lower in the top few recommendations. Another observation is that improvements are not consistent with all datasets. With LFM-50U, for example, bagging only slightly outperforms the baseline ISGD -- and only with $M \geq 32$ --, while with PLC-STR, the improvement is much higher in proportion, even with lower $M$.

It is also clear that the time overheads grow linearly with the number of bootstrap models. However, the overhead in model update times is not very relevant in practice, given that the baseline update times are very low in ISGD -- with $M = 64$ the highest update time falls below 400ms. The overhead at recommendation time is more evident, when aggregating results from the $M$ bootstrap nodes. Fortunately, as with most ensemble techniques, parallel processing can be trivially used to alleviate this overhead. Additionally, there may be room for code optimization or approximate methods that require less and/or more efficient computations.

\section{Conclusions}
\label{sec:conclusions}

Bagging is a an ensemble technique successfully used with many machine learning algorithms, however it has not been thoroughly studied in recommendation problems, and particularly with incremental algorithms. In this paper, we experiment online bagging with an incremental matrix factorization algorithm that learns from unbounded streams of positive-only data. Our results suggest that with manageable overheads, accuracy clearly improves -- more than 35\% in some cases --, especially as the number of recommended items increases. In the near future, we intend to experiment this and other online ensemble methods in a larger number of stream-based recommendation algorithms. 

\section{Acknowledgments}
Project “TEC4Growth – Pervasive Intelligence, Enhancers and Proofs of Concept with Industrial Impact/NORTE-01-0145-FEDER-000020” is financed by the North Portugal Regional Operational Programme (NORTE 2020), under the PORTUGAL 2020 Partnership Agreement, and through the European Regional Development Fund (ERDF). This work is also partially funded by the European Commission through project MAESTRA (Grant no. ICT-2013-612944). We thank Ubbin Labs, Lda. for kindly providing data from Palco Principal.

\bibliographystyle{splncs03}
\bibliography{EPIA17}
\end{document}